\documentclass[12pt]{article}
\pdfoutput=1
\usepackage[utf8]{inputenc}
\usepackage{geometry}
\usepackage{graphicx}
\usepackage{array}
\usepackage{subfig}
\usepackage{slashed}
\usepackage{amsmath}
\usepackage{amsfonts}
\usepackage{amssymb}
\usepackage[cm]{fullpage}
\usepackage{cite}
\usepackage{epsfig}
\usepackage{comment}
\usepackage{hyperref}
\usepackage{feynmp}
\usepackage{cancel}
\usepackage{mathrsfs}

\usepackage{cleveref}
\crefname{section}{§}{§§}
\Crefname{section}{§}{§§}

\numberwithin{equation}{section}

\def\ip{${\mathscr I}^+$}

\def\e{{\epsilon}}
\def\cs{${\cal S}$}

\def\p{\partial}
\def\bz{{\bar z}}

\def\bw{{\bar w}}
 
\def\0{{(0)}}
\def\1{{(1)}}
\def\2{{(2)}}

\def\ci{{\mathscr I}}

\def\<{\langle }
\def\>{\rangle }
\def\bw{{\bar w}}
\def\res{\mathop{\text{Res}}\limits}

\newcommand{\bea}{\begin{eqnarray}}
\newcommand{\eea}{\end{eqnarray}}
\newcommand{\be}{\begin{equation}}
\newcommand{\ee}{\end{equation}}
\newcommand{\ba}{\begin{align}}
\newcommand{\ea}{\end{align}}

\newcommand{\tr}[1]{\text{tr} \left[ #1 \right]}

\def\dim#1{\lbrack\!\lbrack #1 \rbrack\!\rbrack }

   \makeatletter
  \let\over=\@@over \let\overwithdelims=\@@overwithdelims
  \let\atop=\@@atop \let\atopwithdelims=\@@atopwithdelims
  \let\above=\@@above \let\abovewithdelims=\@@abovewithdelims
\renewcommand\section{\@startsection {section}{1}{\z@}%
                                   {-3.5ex \@plus -1ex \@minus -.2ex}%nn
                                   {2.3ex \@plus.2ex}%
                                   {\normalfont\large\bfseries}}

\renewcommand\subsection{\@startsection{subsection}{2}{\z@}%
                                     {-3.25ex\@plus -1ex \@minus -.2ex}%
                                     {1.5ex \@plus .2ex}%
                                     {\normalfont\bfseries}}

\newcommand{\bra}[1]{\left< #1 \right|}
\newcommand{\ket}[1]{\left| #1 \right>}

\newcommand{\beq}{\begin{equation}}
\newcommand{\eeq}{\end{equation}}
\newcommand{\beqa}{\begin{eqnarray}}
\newcommand{\eeqa}{\end{eqnarray}}
\newcommand{\beqar}{\begin{eqnarray*}}

\newcommand{\Or}{\mathcal{O}}

\newcommand{\cl}{{\cal L}}

\newcommand{\cm}{{\cal M}}
\newcommand{\ve}{{\varepsilon}}

\newcommand{\F}{{\cal F}}
\newcommand{\A}{{\cal A}}
\def\[{\[}
\def\]{\]}

\newcommand{\C}{{\cal C}}

\def\mcc{{\mathbb C}}
\def\mrr{{\mathbb R}}

\newcommand{\bd}[1]{\begin{fmffile}{#1}\begin{fmfgraph*}}
\newcommand{\ed}{\end{fmfgraph*}\end{fmffile}}

\begin{document}

\begin{titlepage}
\unitlength = 1mm~\\
\vskip 3cm
\begin{center}

{\LARGE{\textsc{2D Kac-Moody Symmetry of 4D Yang-Mills Theory}}}

\vspace{0.8cm}
Temple He, Prahar Mitra and Andrew Strominger

\vspace{1cm}

{\it  Center for the Fundamental Laws of Nature, Harvard University,\\
Cambridge, MA 02138, USA}

\vspace{0.8cm}

\begin{abstract}
Scattering amplitudes of any four-dimensional theory with nonabelian gauge group $\mathcal G$ may be recast as two-dimensional correlation functions on the asymptotic two-sphere at null infinity. The soft gluon theorem is shown, for massless theories at the semiclassical level, to be the Ward identity of a holomorphic two-dimensional $\mathcal G$-Kac-Moody symmetry  acting on these correlation functions. Holomorphic Kac-Moody current insertions are positive helicity soft gluon insertions. The Kac-Moody transformations are a $CPT$ invariant subgroup of gauge transformations which act nontrivially at null infinity and comprise the four-dimensional asymptotic symmetry group. 

\end{abstract}

\vspace{1.0cm}

\end{center}

\end{titlepage}

\pagestyle{empty}
\pagestyle{plain}

\def\vx{{\vec x}}
\def\p{\partial}
\def\po{$\cal P_O$}

\pagenumbering{arabic}

\tableofcontents
\bibliographystyle{utphys}

\section{Introduction}

The $n$-particle scattering amplitudes $\A_n$ of any four-dimensional quantum field theory (QFT$_4$) can be described as a collection of $n$-point correlation functions on the two-sphere ($S^2$) with coordinates $(z,\bz)$
\be\label{dyu}
\A_n=\<O_1(E_1,z_1,\bz_1)\cdots O_n(E_n,z_n,\bz_n)\>, 
\ee
where $O_k$ creates (if $E_k < 0$) or annihilates (if $E_k > 0$) an asymptotic particle with energy $|E_k|$ at the point $(z_k, \bz_k)$ where the particle crosses the asymptotic $S^2$ at null infinity ($\ci$). The alternate description \eqref{dyu} is obtained from the usual momentum space description by simply trading the three independent components of the on-shell four momentum $p_k^\mu$ (subject to $p_k^2=-m_k^2$) with the three quantities $(E_k,z_k,\bz_k)$. 

The Lorentz group $SL(2,\mcc) $ acts as the global conformal group on the asymptotic $S^2$ according to 
\be\label{lsy} 
z\to {az+b \over cz+d} , 
\ee
with $ad-bc=1$. Hence, in this respect, \textit{Minkowskian} QFT$_4$ amplitudes resemble \textit{Euclidean} two-dimensional conformal field theory (CFT$_2$) correlators. It is natural to ask what other properties  QFT$_4$ scattering amplitudes, expressed in the form \eqref{dyu}, have in common with conventional CFT$_2$ correlators, and more generally whether a holographic relation of the form \textit{Minkowskian QFT$_4 = $ Euclidean CFT$_2$} might plausibly exist when gravity is included.\footnote{The results of \cite{Cachazo:2014fwa, Banks:2003vp, deBoer:2003vf,barnichtroessaert, Kapec:2014opa} suggest that for quantum gravity scattering amplitudes the $SL(2,\mcc)$ Lorentz symmetry \eqref{lsy} is enhanced to the infinite-dimensional local 2D conformal symmetry.} In this paper, we consider tree-level scattering of massless particles in 4D nonabelian gauge theories with gauge group $\cal G$.  A salient feature of all such amplitudes is that soft gluon scattering is controlled by the soft gluon theorem \cite{Berends:1988zn}. A prescription is given for completing the hard \cs-matrix (in which all external states have $E_k \neq 0$) to an \cs-correlator which includes positive helicity soft gluons at strictly zero energy.  It is shown that the content of the soft gluon theorem at tree-level is that the positive helicity soft gluon insertions are holomorphic 2D currents which generate a 2D $\cal G$-Kac-Moody algebra in the \cs-correlator! Turning the argument around, the soft gluon theorem can be derived as a tree-level Ward identity of the Kac-Moody symmetry.\footnote{A similar Kac-Moody algebra was studied in \cite{nair} in the context of MHV amplitudes.}

Moreover, we show that the Kac-Moody symmetries are equivalent to the asymptotic symmetries of 4D gauge theories described in \cite{as}. They are $CPT$-invariant gauge transformations, which are independent of advanced or retarded time and take angle-dependent values on $\ci$. $CPT$ invariance requires that the gauge transformation at any point on $\ci^+$ equals that at the $PT$ antipode on $\ci^-$. Such transformations act nontrivially on the asymptotic physical states and comprise the asymptotic symmetry group. These are the gauge theory analogs of BMS transformations in asymptotically flat gravity \cite{bms,barnichtroessaert,ashtekar,as1,hms,Kapec:2014opa}. The abelian $U(1)$ case was discussed in \cite{hmps,Lysov:2014csa,Kapec:2014zla} and related recent discussions of symmetries, infrared divergences and soft theorems are in \cite{Cachazo:2013hca,Cachazo:2014fwa,Casali:2014xpa,Schwab:2014xua,Bern:2014oka, He:2014bga,larkoski, Larkoski:2014bxa, Afkhami-Jeddi:2014fia, Adamo:2014yya,Geyer:2014lca,schwab, Bianchi:2014gla, Balachandran:2014hra, Broedel:2014fsa, Bern:2014vva, Campoleoni:2014tfa, Mohd:2014oja, Lambert:2014poa,Campiglia:2015yka}.

The asymptotic symmetries of gravity (or QED) are spontaneously broken in the perturbative vacuum and the soft gravitons (or photons) were shown to be the resulting Goldstone bosons \cite{as1,hms,hmps}. Analogously, the standard rules of Yang-Mills perturbation theory presume a trivial flat color frame on $\ci$. In this paper we see that this trivial frame is not invariant under the non-constant Kac-Moody transformations and the large gauge symmetry is spontaneously broken, with the soft gluons being the corresponding Goldstone bosons. 

The nonabelian interactions of Yang-Mills theory lead to some surprising new features that are not present in parallel analyses of gravity \cite{as1,hms} and QED \cite{hmps}.  As pointed out to us by S. Caron-Huot \cite{schuot,jmald}, the double-soft limit of the \cs-matrix involving one positive and one negative helicity gluon is ambiguous. The result depends on the order in which the gluons are taken to be soft. Hence a prescription must be given for defining the double-soft boundary of the \cs-matrix.  We adopt the prescription that positive helicities are always taken soft first. With this prescription there is one holomorphic $\cal G$-Kac-Moody from positive helicity soft gluons, but not a second one from negative helicity soft gluons. 

The soft gluon theorem has well-understood  universal corrections due to IR divergences which appear only at one loop, see e.g. \cite{mattilya}.  These will certainly affect any extension of the present discussion beyond tree-level.  Since an infinite number of relations among \cs-matrix elements remain, an asymptotic symmetry may survive these corrections. However it is not clear if it can still be understood as a Kac-Moody symmetry. Corrections do $not$ appear at the level of the integrands studied in the amplitudes program \cite{He:2014bga,Bern:2014oka,Cachazo:2014dia,Elvang:2013cua} or in contexts requiring the soft limit to be taken prior to the removal of the IR regulator \cite{Cachazo:2014dia, Bern:2014oka,mattilya}. Hence the Kac-Moody symmetry is relevant in some contexts to all loops. We leave this issue, as well as the generalization to massive particle scattering, to future investigations.  

This paper is organized as follows. Section \ref{conventions} establishes our notation and conventions. In section \ref{asymptoticfields}, we introduce the various asymptotic fields used in the paper and discuss the asymptotic symmetries of nonabelian gauge theories. In section \ref{kacmoody}, we show that the soft gluon theorem is the Ward identity of  a holomorphic Kac-Moody symmetry which can also be understood as an asymptotic gauge symmetry.  In section \ref{antiholcurr}, we show that the double-soft ambiguity of the \cs-matrix obstructs the appearance of a second antiholomorphic Kac-Moody. Finally, section \ref{flatconn} contains a  preliminary discussion of Wilson line insertions, SCET fields and an operator realization of the flat gauge connection on $\ci$.  

\section{Conventions and notation}\label{conventions}
We consider a nonabelian gauge theory with group ${\cal G}$ and associated Lie algebra ${\mathfrak g}$. Elements of ${\cal G}$ in representation $R_k$ are denoted by $g_k$, where $k$ labels the representation. The corresponding hermitian generators of ${\mathfrak g}$ obey 
\begin{equation}
\begin{split}
\left[ T_k^a , T_k^b \right] = i f^{abc} T_k^c, 
\end{split}
\end{equation}
where  $a = 1, \cdots, \dim {\mathfrak g}$ and the sum over repeated Lie algebra indices is implied. The adjoint elements of ${\cal G}$ and generators of ${\mathfrak g}$ are denoted by $g$ and $T^a$ respectively with $(T^a)_{bc} = - i f^{abc}$. The real antisymmetric structure constants $f^{abc}$ are normalized so that 
\begin{equation}
\begin{split}\label{fnorm}
 f^{acd} f^{bcd} = \delta^{ab} =\tr{T^aT^b}.
\end{split}
\end{equation}
The four-dimensional matrix valued gauge field is $\A_\mu = \A_\mu^a T^a$, where a  $\mu$ index here and hereafter refers to flat  Minkowski coordinates in which the metric is 
\begin{equation}\label{mm}
\begin{split}
	ds^2 = -dt^2 + d\vec{x} \cdot d \vec{x} ,
\end{split}
\end{equation}
with $\vec{x} = (x^1, x^2, x^3)$ satisfying $\vec{x} \cdot \vec{x} = r^2$. 
We also use  retarded coordinates
\begin{equation}\label{retardedcoord}
	ds^2 = -du^2 - 2du dr + 2r^2\gamma_{z\bz}dz d\bz,
\end{equation}
where $u = t - r$ and $\gamma_{z\bz} = \frac{2}{(1+z\bz)^2}$ is 
the round metric on the sphere.   $\ci^+$ is the null $S^2 \times {\mathbb R}$ boundary at $r = \infty$ with coordinates $(u,z,\bz)$. It has boundaries at $u = \pm \infty$, which we denote $\ci^+_\pm$.

$PT$-conjugate advanced coordinates are
\begin{equation}\label{retdedcoord}
	ds^2 = -dv^2 +2dvdr + 2r^2\gamma_{z\bz}dz d\bz,
\end{equation}
where $v=t+r$. $\ci^-$ is the null $S^2 \times \mrr$ boundary at $r = \infty$ with coordinates $(v,z,\bz)$. It has boundaries at $v = \pm \infty$, which we denote as $\ci^-_\pm$.  

Advanced and retarded coordinates are related to the flat coordinates in \eqref{mm} by 
\begin{equation}
\begin{split}
	t = v - r, \quad x^1 + i x^2 = -\frac{2rz}{1+z\bz}, \quad x^3 =- \frac{r \left( 1 - z \bz \right)}{1 + z\bz },
\end{split}
\end{equation}
and \begin{equation}
\begin{split}
	t = u + r, \quad x^1 + i x^2 = \frac{2rz}{1+z\bz}, \quad x^3 = \frac{r \left( 1 - z \bz \right)}{1 + z\bz }.
\end{split}
\end{equation}
In particular, note that the point with coordinates $(r,v, z,\bz)$ in advanced coordinates is antipodally related by $PT$ to the point with coordinates $(r, u,z,\bz)$ in retarded coordinates. 

The field strength corresponding to $\A_\mu$ is 
\begin{equation}
\begin{split}
\F_{\mu\nu} = \p_\mu \A_\nu - \p_\nu \A_\mu - i [ \A_\mu, \A_\nu ] = \F_{\mu\nu}^a T^a . 
\end{split}
\end{equation}

The theory is invariant under gauge transformations
\be
\begin{split}\label{gauge_trans}
	\A_\mu &\to g \A_\mu g^{-1}  + i g \p_\mu g^{-1}, \\
	\phi_k &\to g_k \phi_k, \\
 	j_\mu^M &\to g j_\mu^M g^{-1},
\end{split}
\ee
where $\phi_k$ are matter fields in representation $R_k$ and $j_\mu^M$ is the matter current that couples to the gauge field. The infinitesimal gauge transformations with respect to ${\hat \ve} = {\hat \ve}^a T^a$ (where $g = e^{ i {\hat \ve} } $) are
\begin{equation}
\begin{split}
\delta_{\hat \ve} \A_\mu &= \p_\mu {\hat \ve} - i [ \A_\mu , {\hat \ve} ]  , \\
\delta_{\hat \ve} \phi_k &= i {\hat \ve}^a T^a_k  \phi_k , \\
\delta_{\hat \ve} j_\mu^M &=  - i [ j_\mu^M , {\hat \ve} ] . 
\end{split}
\end{equation}

The bulk equations that govern the dynamics of the gauge field are
\be
	\nabla^\nu \F_{\nu\mu} - i \left[ \A^\nu, \F_{\nu\mu} \right] = g_{YM}^2 j^M_\mu,
\ee
where $\nabla^\mu$ is the covariant derivative with respect to the spacetime metric.

In this paper, we study massless scattering amplitudes. Following \cite{hmps}, we find it convenient to parametrize massless momenta $p^2 = 0$ as
\begin{equation}
\begin{split}\label{momentumparametrization}
p^\mu = \frac{\omega}{1+z\bz} \left( 1 + z \bz , z + \bz , - i  ( z - \bz ) , 1 - z \bz  \right) . 
\end{split}
\end{equation}
For simplicity, we also denote $\vec{p} = \omega {\hat x}$ where
\begin{equation}
\begin{split}\label{xhatdef}
{\hat x} = \frac{1}{1+z\bz} \left(   z + \bz , - i  ( z - \bz ) , 1 - z \bz  \right) . 
\end{split}
\end{equation}

\section{Asymptotic fields and symmetries}\label{asymptoticfields}

In this section, we give our conventions for the asymptotic expansion around $\ci$ (see \cite{as} for more details), specify the gauge conditions and boundary conditions, and describe the residual large gauge symmetry. 

We work in temporal gauge
\begin{equation}
\begin{split}\label{gaugecond} 
\A_u &=0 . 
\end{split}
\end{equation}
In this gauge, we can expand the gauge fields near $\ci^+$ as
\be\begin{split}
	\mathcal A_z(r,u,z,\bz) &= A_z(u,z,\bz) + \mathcal O\left( 1 /r \right) , \\
	\mathcal A_r(r,u,z,\bz) &= \frac{1}{r^2}A_r(u,z,\bz) + \mathcal O\left( 1 / r^3 \right),
\end{split}\ee
where the leading behavior of the gauge field is chosen so that the charge and energy flux through $\ci^+$ is finite. The full four-dimensional gauge field is determined by the equations of motion in terms of $A_z(u)$, which forms the boundary data of the theory. 

The leading behavior of the field strength is $\F_{ur} = \Or(1/r^2)$ and $\F_{uz},\F_{z\bz} = \Or(1)$ with leading coefficients
\be\begin{split}\label{field_str}
	F_{ur} &= \p_u A_r , \\
	F_{uz} &= \p_u A_z , \\
	F_{z\bz} &= \p_z A_\bz - \p_{\bz}A_z - i [A_z, A_\bz]. \\
\end{split}\ee
We will be interested in configurations that revert to the vacuum in the far future, i.e. 
\begin{align}
F_{ur} |_{\ci^+_+} = F_{uz} |_{\ci^+_+} = F_{z\bz} |_{\ci^+_+} =  0 . \label{flatnesscond} 
\end{align}
\eqref{flatnesscond} implies
\begin{equation}
\begin{split}\label{uzflatdef}
{\cal U}_z \equiv A_z  |_{\ci^+_+} = i {\cal U} \p_z {\cal U}^{-1} , 
\end{split}
\end{equation}
where ${\cal U}(z,\bz) \in {\cal G}$.
A residual gauge freedom near $\ci^+$ is generated by an arbitrary function $\ve(z,\bz)$ on the asymptotic $S^2$. These create zero-momentum gluons and will be referred to as large gauge transformations.  Under finite large gauge transformations ${\cal U} \to g {\cal U}$. We also define the soft gluon operator
\begin{equation}
\begin{split}
N_z &\equiv \int_{-\infty}^\infty du F_{uz}   = {\cal U}_z - A_z |_{\ci^+_-}  . 
\end{split}
\end{equation}

Near $\ci^-$, the temporal gauge condition implies
\begin{align}
\A_v = 0 . 
\end{align}
We expand the gauge fields as $\A_z  = B_z + \Or(r^{-1})$, $\A_r = \frac{1}{r^2} B_r + \Or(r^{-3})$. The field strength has leading behaviour $\F_{vr} \sim \Or(1/r^2)$ and $\F_{vz} , \F_{z\bz} = \Or(1)$ with leading coefficients
\be\begin{split}
	G_{vr} &= \p_v B_r , \\
	G_{vz} &= \p_v B_z , \\
	G_{z\bz} &= \p_z B_\bz - \p_{\bz}B_z - i [B_z, B_\bz]. \\
\end{split}\ee
Configurations that begin from the vacuum in the far past satisfy
\begin{equation}
\begin{split}\label{flatnesscond1}
G_{ur} |_{\ci^-_-}  = G_{vz} |_{\ci^-_-}  = G_{z\bz} |_{\ci^-_-}  = 0  . 
\end{split}
\end{equation}
The four-dimensional gauge field is uniquely determined by the boundary data $B_z(v)$. 

Residual gauge freedom near $\ci^-$ is generated by an arbitrary function $\ve^-(z,\bz)$ on the asymptotic $S^2$. Furthermore, \eqref{flatnesscond1} implies
\begin{equation}
\begin{split}
{\cal V}_z \equiv B_z |_{\ci^-_-} = i {\cal V} \p_z {\cal V}^{-1}  , 
\end{split}
\end{equation}
On $\ci^-$, we define the soft gluon operator
\begin{equation}
\begin{split}
M_z &\equiv \int_{-\infty}^\infty dv G_{vz}  = B_z |_{\ci^-_+} - {\cal V}_z    . 
\end{split}
\end{equation}

The classical scattering problem, i.e. to determine the final data $A_z(u)$ given a set of initial data $B_z(v)$ is defined only up to the large gauge transformations generated by both $\ve$ and $\ve^-$ that act separately on the initial and final data. Clearly, there can be no sensible scattering problem without imposing some relation between $\ve$ and $\ve^-$. To do this, we match the gauge field at $i^0$. Lorentz invariant matching conditions are
\begin{equation}
\begin{split}\label{matchcond}
A_z |_{\ci^+_-} = B_z |_{\ci^-_+}  .
\end{split}
\end{equation}
This is preserved by 
\begin{equation}
\begin{split}\label{gaugecondasdsa}
\ve(z,\bz) = \ve^-(z,\bz) . 
\end{split}
\end{equation}
Note that because of the antipodal identification of the null generators of $\ci^\pm$ across $i^0$, the gauge parameter $\ve(z,\bz)$ is \textit{not} the limit of a function that depends on the angle in Minkowskian $(t,r)$ coordinates. Rather, it goes to the same value at the beginning and end of light rays crossing through the origin of Minkowski space. $\ve$ is then a Lie algebra valued function (or section) on the space of null generators of $\ci$.

\section{Holomorphic  soft gluon current}\label{kacmoody}

In this section, we show that the soft theorem for outgoing positive helicity gluons (or equivalently incoming negative helicity gluons) is the Ward identity of the 
holomorphic large gauge transformations and takes the form of a holomorphic ${\cal G}$-Kac-Moody symmetry acting on the $S^2$ on $\ci$.

Let $O_k(E_k, z_k,\bz_k)$ denote an operator which creates or annihilates a colored hard particle with energy $E_k\neq 0$ crossing the $S^2$ on $\ci$ at the point $z_k$.\footnote{For instance, for scalar particles
\begin{equation}
\begin{split}
O_k (E_k , z_k ,\bz_k) = - \frac{4\pi}{E_k} \int_{-\infty}^\infty d u e^{i E_k u  } \p_u \lim_{r \to \infty} \left[  r \phi_k ( u , r , z_k , \bz_k )  \right]  . \nonumber
\end{split}
\end{equation}
}
We denote the standard $n$-particle hard amplitudes by 
\be\A_n(z_1,\ldots ,z_n)=\<O_1\cdots O_n\>_{U=1}.\ee
There are no traces here, so $\A_n$ has $n$ suppressed color indices. Since the gauge field vanishes at infinity, the asymptotic $S^2$ has a flat connection $U_z = i U \p_z U^{-1}$, where $U \in {\cal G}$.\footnote{$U_z$ should not be confused with ${\cal U}_z$ (defined in \eqref{uzflatdef}).} In order to compare the color of particles emerging at different points on the $S^2$, this connection must be specified. The $U=1$ subscript here indicates the fact that the standard perturbation theory presumes the trivial connection $U_z = 0$.\footnote{For $U_z=0$  an outgoing  configuration with a red quark at the north pole and a red bar quark at the south pole is a color singlet state which can be created by a colorless incoming state. For more general choices of $U_z$ this will not be the case.}

The hard \cs-matrix has soft boundaries where gluon momenta vanish. We wish to give a prescription to extend, or `compactify' the \cs-matrix to a larger object that includes these boundaries. Since zero-energy gluons are not obviously either incoming or outgoing, the \cs-matrix so compactified is not obviously a matrix mapping $in$ states to $out$ states.  Hence we will refer to the compactified \cs-matrix as the \cs-correlator.\footnote{In the abelian examples of gravity and QED \cite{as1,hms,Kapec:2014opa,hmps}, it is possible to view the \cs-correlator as a conventional \cs-matrix. However, the noncommutativity (see \eqref{slm}) of the multi-gluon soft limits persists even if one gluon  is outgoing ($q^0 > 0$) and the other incoming ($q'^0 < 0$). This means that the soft limit on an $out$ state does not commute with the soft limit on an $in$ state, creating difficulties for the reinterpretation of the \cs-correlator as an \cs-matrix.}

\subsection{Soft gluon theorem} 

In this section, we will show that insertions of the soft gluon current $J_z$, defined by 
\begin{equation}
\begin{split}\label{jzdef}
J_z \equiv - \frac{4\pi}{g_{YM}^2} \left( N_z - M_z \right) = \frac{4\pi}{g_{YM}^2} \left( \int dv G_{vz} - \int du F_{uz} \right), 
\end{split}
\end{equation}
into the hard tree-level \cs-matrix are determined by the soft gluon theorem. In its conventional momentum space form, this theorem states (see appendix \ref{appa})
\be \label{tok} 
 \<O_1(p_1)\cdots O_n(p_n); O^a(q,  \e)\>_{U=1}= g_{YM} \sum_{k=1}^n{p_k \cdot \e \over p_k \cdot q}\<O_1(p_1)\cdots T_k^aO_k(p_k)\cdots O_n(p_n)\>_{U=1} +\Or (q^0), 
\ee
where $O^a(q,\e)=\tr{T^a O(q,\e)}$ creates or annihilates, depending on the sign of $q^0$, a soft gluon with momentum $\vec{q}$ and polarization $\e^\mu$, and $T^a_k$ is a generator in the representation carried by $O_k$. Gauge invariance of the theory requires that the right hand side vanishes when $\e = q$. This implies
\begin{equation}
\begin{split}\label{chargeconservation}
\sum_{k=1}^n \<O_1(p_1)\cdots T_k^aO_k(p_k)\cdots O_n(p_n)\>_{U=1} = 0, 
\end{split}
\end{equation}
which is global color conservation. Using the notation of our present paper and assuming $\e \neq q$, for a positive helicity gluon with massless particles ($p_k^2=0$), \eqref{tok} becomes 
\be\label{ftip}\<J^a_zO_1\cdots O_n\>_{U=1}=\sum_{k=1}^n{1\over z-z_k} \<O_1\cdots T^a_kO_k\cdots O_n\>_{U=1} ,\ee
where $J^a_z \equiv \tr{T^a J_z}$.  This was shown in \cite{maldacena,as} and is reviewed in the appendix. The collinear $q\cdot p_k \to 0$ singularities of \eqref{tok} become the poles at $z=z_k$ in \eqref{ftip}. The soft pole in \eqref{tok} is absent in \eqref{ftip} simply because the definition of $J^a_z$ involves the zero mode of the field strength rather than the gauge field and hence an extra factor of the soft energy. 

\subsection{Kac-Moody symmetry}

Since $\p_\bz J_z=0$ away from operator insertions, $J_z$ is a holomorphic current. Consider a contour $\C$ and an infinitesimal gauge transformation $\ve^a(z)$ which is holomorphic ($\p_\bz \ve^a =0$) inside $\C$. It follows from \eqref{ftip} that 
\bea\label{jjgj} \<J_\C(\ve)O_1\cdots O_n\>_{U=1}=\sum_{k\in \C} \<O_1\cdots \ve_k(z_k)O_k\cdots O_n\>_{U=1} ,\eea
where $\ve_k(z_k)=\ve^a(z_k)T^a_k$ and 
\be J_\C(\ve)\equiv \oint_\C \frac{dz}{2\pi i} \tr{ \ve  J_z } , \ee
and the sum $k\in \C$ includes all insertions inside the contour $\C$.  Moreover from the soft theorem with multiple $J_z$ insertions one finds
 \bea\label{jj}\<J_\C(\ve)J_wO_1\cdots O_n\>_{U=1}=\sum_{k\in \C} \<J_w O_1\cdots \ve_k(z_k)O_k\cdots O_n\>_{U=1} +  \<\ve(w)J_wO_1 \cdots O_n\>_{U=1} ,\eea
where the last term is added only when $w$ is also inside $\C$. 
 
\eqref{jj} is a very familiar formula in two-dimensional conformal field theory. It is the Ward identity of a holomorphic  Kac-Moody symmetry for the group ${\cal G}$. The absence of a term with no $J_w$ on the right hand side of \eqref{jj} indicates that the Kac-Moody level is zero (at tree-level). Hence the \cs-correlators for any massless theory with nonabelian gauge group ${\cal G}$ transform under a holomorphic level-zero ${\cal G}$-Kac-Moody action!

\subsection{Asymptotic symmetries} 

In this subsection, the Kac-Moody symmetry is  identified with holomorphic large gauge symmetry of the gauge theory. According to  \eqref{gauge_trans} under the action of the asymptotic symmetry transformation $U$
\begin{equation}
\begin{split}
O_k(z_k,\bz_k) \to U_k (z_k,\bz_k) O_k(z_k,\bz_k),
\end{split}
\end{equation}
where $U_k$ acts in the representation of $O_k$. \cs-correlators for general $U$ are simply related to those for $U=1$ 
\be\label{pkj}
\< J^a_z O^{i_1}_1\cdots\>_{U}= U(z,\bz)^{ab} U_1 (z_1,\bz_1)^{i_1j_1} \cdots\< J^b_z O^{j_1}_1\cdots \>_{U=1}.
\ee 

To compare the asymptotic symmetry action \eqref{pkj} with the Kac-Moody action \eqref{jjgj}, consider infinitesimal complexified transformations of the form 
\be U(z,\bz) =1+i \ve(z) +\cdots , \ee
which are holomorphic inside the contour $\C$ and vanish outside. In that case \eqref{pkj} linearizes to 
\be \delta_\ve \<O_1\cdots O_n\>_{U=1}=  i \sum_{k\in \C} \<O_1\cdots \ve_k(z_k)O_k\cdots O_n\>_{U=1}, \ee
where the operator insertions could also include a postive-helicity soft gluon. Comparing with 
\eqref{jjgj} we see that 
\be\label{ut} - i \delta_\ve \<O_1\cdots O_n\>_{U=1}= \<J_\C(\ve)O_1\cdots O_n\>_{U=1}. \ee
Hence, $J_\C(\e)$  generates holomorphic asymptotic symmetry transformations
\be \label{jgu} J_\C(\ve)=\int_{D_C} d^2 z \gamma_{z\bz}\ve^a{\delta \over \delta U^a} , \ee
where $D_C$ is the region inside ${\cal C}$ and $U^a$ is the Lie algebra element corresponding to $U$, that is, $U = e^{iU^aT^a}$.

Let $\C_0$ be any contour that divides the incoming and outgoing particles. For $\ve$ holomorphic on the incoming side of $\C_0$, the corresponding $J_{\C_0}(\ve)$  is then the charge that generates the asymptotic symmetries on the incoming state. If $\ve$ is holomorphic and non-constant on the incoming side of $\C_0$, it extends to a meromorphic section which must have poles on the outgoing side whose locations we denote $w_1,\ldots,w_p$. We may also evaluate the contour integral by pulling it over the outgoing state.
Equating this with \eqref{jjgj} one finds, for any meromorphic section $\ve$
\be  - i \delta_\ve \<O_1\cdots O_n\>_{U=1}= -\sum_{i=1}^p \<\tr{\ve J_w}_{w_i}O_1\cdots O_n\>_{U=1} , \ee
where
\begin{equation}
\begin{split}
\tr{\ve J_w}_{w_i} = \res_{w \to w_i} \tr{ \ve J_w }  . 
\end{split}
\end{equation}
This is another form of the soft gluon theorem. It states that \cs-correlators are invariant under the asymptotic symmetries  up to insertions of the soft gluon current. The appearance of the  inhomogenous term on the right hand side implies that the $U=1$ vacuum spontaneously breaks the symmetry. The soft gluons are the associated Goldstone bosons. 
Indeed, when $p = 0$, i.e. when $\ve$ is a globally holomorphic function on the sphere (and therefore a constant), we have
\begin{equation}
\begin{split}
\delta_\ve \<O_1\cdots O_n\>_{U=1} = 0  , 
\end{split}
\end{equation}
which is precisely \eqref{chargeconservation}. This indicates that the subgroup of constant global asymptotic color rotations is not spontaneously broken, as expected. 

One might think that the Kac-Moody symmetry does not capture {\it all} of the asymptotic symmetry group, since the transformations are restricted to be holomorphic within some contour $\C$. However, this is an irrelevant restriction.  The \cs-correlator identities depend only on the $n$  values $\ve_k=\ve(z_k)$ of $\ve$ at the $n$ operator insertions. For $any$ choice of  $\ve_k$  there exists a holomorphic $\ve(z)$ inside some $\C$ such that $\ve(z_k)=\ve_k$ at the positions of operator insertions. Hence the holomorphicity does not preclude consideration of any gauge transformation on Fock space states, and all nontrivial relations among \cs-correlation functions can be derived from the Kac-Moody symmetry.  In particular the soft gluon theorem \eqref{ftip} is itself a Ward identity of the the Kac-Moody symmetry.

\section{Antiholomorphic current}\label{antiholcurr}

We have seen that  positive helicity soft gluon currents  $J_z$ generate a holomorphic Kac-Moody symmetry. Naively one might expect that negative helicity soft gluon  currents  $J^a_\bz$ generate a second  Kac-Moody symmetry which is antiholomorphic. This turns out $not$ to be the case 
for a very interesting reason. 

The crucial observation is due to \cite{jmald,schuot}.  Consider a boundary of the \cs-matrix near which two gluons become soft. One finds
\be 
\begin{split}
\A_{n+2}(p_1,\ldots, p_n;q,\e,a;q',\e',b) &=  g_{YM}^2 \sum_{k=1}^n{\e\cdot p_k \over q\cdot p_k}\sum_{j=1}^n{\e'\cdot p_j \over q'\cdot p_j} \<O_1\cdots T^a_kO_k\cdots T^{b}_jO_j\cdots O_n\>_{U=1} \\
&\quad  - i g_{YM}^2 f^{abc}\sum_{j=1}^n \frac{\e' \cdot p_j}{q'\cdot p_j}\frac{\e \cdot q'}{q \cdot q'} \<O_1 \cdots T_j^c O_j \cdots O_n \>_{U=1} + \Or(q^0,q'^0) ,
\end{split}
\ee
where the above limit has been computed by taking $q \to 0$ first. Surprisingly, the right hand side actually depends on the order of limits and
\be \begin{split}
	\label{slm} \left[\lim_{q\to 0},\lim_{q'\to 0}\right]\A_{n+2}&(p_1,\ldots, p_n;q,\e,a;q',\e',b) = i g_{YM}^2 f^{abc}\sum_{k=1}^n \left(\frac{\e \cdot p_k}{p_k \cdot q}   - \frac{\e \cdot q'}{q \cdot q'}  \right) \left( \frac{\e' \cdot p_k}{q' \cdot p_k} -  \frac{\e' \cdot q}{q \cdot q'} \right) \\
	&~~~~~~~~~~~~~~~~~~~~~~~~~~~~~~~~~~~~~~~~~~~~~~~~\times  \<O_1\cdots T^a_kO_k \cdots O_n\>_{U=1}+\Or\left(q^0,q'^0\right).
\end{split}
\ee
In the special case that the helicities are the same, then the right hand side of the above expression vanishes and the limits commute. In this case, the \cs-matrix can be extended to its soft boundaries unambiguously. When the helicities are not the same, the value of the \cs-matrix at the soft boundary is ambiguous. In terms of currents, taking the positive helicity gluon to zero first gives 
\be\label{dsv} J_z^aJ_\bw^b\sim - {i f^{abc}  \over z-w}J_\bw^c,\ee
while in the other order we have
\be J_z^aJ_\bw^b\sim- {i f^{abc}  \over \bz-\bw}J_z^c.\ee 

Thus, the extension (or `compactification') of  the \cs-matrix to all soft boundaries requires a prescription. In this paper we adopt the prescription that positive helicity gluon momenta are always taken to zero \textit{before} negative helicity gluon momenta. With this prescription, it follows from \eqref{dsv} that the current $J^a_z$ generates a Kac-Moody symmetry, under which $J^a_\bz$ transforms in the adjoint. $J^a_\bz$ itself does not generate a symmetry. A prescription which treats $J^a_z$ and $J^a_\bz$ symmetrically yields no symmetry, while taking negative helicity momenta to zero first gives one antiholomorphic Kac-Moody symmetry generated by $J_\bz^a$.

The situation is reminiscent of three-dimensional Chern-Simon gauge theory on a manifold with a boundary parameterized by $(z,\bz)$.  A priori, one might have expected $A_z$ and $A_\bz$ to  generate both holomorphic and antiholomorphic ${\cal G}$-Kac-Moody symmetries. However a more careful analysis reveals that boundary conditions must be chosen to eliminate one or the other. Indeed, this may be more than an analogy. The current $J^a_z$ has no time dependence and lives on the $S^2$ at the boundary of the 3-manifold $\ci$, and the addition of a $\theta F\wedge F$ term to the 4D gauge theory action  induces a Chern-Simons term on $\ci$. It would be interesting to understand how such a term affects the present analysis.

\section{Wilson lines and the flat connection on $\ci$}\label{flatconn}

Other types of \cs-correlator insertions besides soft gluon currents are of physical interest and have been considered in the literature. 
This section contains preliminary observations on a few such insertions.

Consider the Wilson line operator 
\be
W_C(u, z_1,z_2)=P \exp \left( i\int_C dx^\mu \A_\mu \right),
\ee
 where $P$ denotes path-ordering and the contour $C$ is chosen such that it initially enters \ip\ at 
$(u,z_1,\bz_1)$ and leaves at $(u,z_2,\bz_2)$ along null lines of varying $r$ and fixed $(u,z,\bz)$. Under holomorphic large gauge transformations
\be W_C(u, z_1,z_2)\to g(z_1)W_C(u, z_1,z_2)g(z_2)^{-1},\ee
where $g(z) \in {\cal G}$. Insertions of $J_z$ in the presence of the Wilson lines are given by the soft theorem\footnote{See $\S$36.3.2 of \cite{Schwartz:2013pla} for details.}
\begin{equation}
\begin{split}
\label{ftp}\<J^a_zW_C(u, z_1,z_2) \cdots \>_{U=1}& = {1\over z-z_1} \<T^aW_C(u, z_1,z_2) \cdots  \>_{U=1}   - {1\over z-z_2} \<W_C(u, z_1,z_2)T^a \cdots \>_{U=1}  + \cdots. 
\end{split}
\end{equation}
From this, we can construct 
\be A_z(u,z,\bz)= - i \lim_{z'\to z}\p_{z} W_C(u, z,z'), \ee
where we take $\C$ to be a short contour from $z'\to z$. It follows from \eqref{ftp}
\be\label{jzaw2pt}\<J^a_zA^b_w O_1\cdots\>_{U=1}= - {i \delta^{ab}\over (z-w)^2} \<O_1\cdots\> - { i f^{abc}\over z-w}\<A^c_wO_1\cdots\> + \cdots .\ee
Hence the action of $J_z$ indeed transforms $A_z$ as a connection on $\ci$ as expected. 
A similar discussion applies to fields on $\ci^-$. 

Recall that $J_z$ was constructed from zero modes of the past and future field strengths (see \eqref{jzdef}). However, $A_z(u)$ has an inhomogeneous term in its gauge transformation and has a soft $u$-independent piece that cannot be constructed from $J_z$. To see this, we expand on $\ci^+$
\be A_z(u, z, \bz)=  \int_{-\infty}^\infty \frac{d\omega}{2\pi} e^{- i\omega u} A^\omega_z(z,\bz) + C_z  , \ee
where
\begin{equation}
\begin{split}\label{Czdef}
C_z \equiv \frac{1}{2} \left( A_z |_{\ci^+_+} + A_z |_{\ci^+_-} \right) = {\cal U}_z - \frac{1}{2} N_z . 
\end{split}
\end{equation}
 Here we have used the fact that functions whose boundary values at $\pm \infty$ do not sum to zero do not have a Fourier transform given in terms of ordinary functions. Radiative insertions in an \cs-matrix involve $A_z^\omega$ and 
\begin{equation}
\begin{split}\label{nndeff}
N_z = - \frac{i}{2} \lim_{\omega \to 0^+} \left(  \omega  A^\omega_z -  \omega  A^{-\omega}_z  \right) . 
\end{split}
\end{equation}

Under a large gauge transformation,
\be \delta_\ve A^\omega_z= - i [ A^\omega_z , \ve],~~~\delta_\ve {\cal U}_z  = \p_z \ve - i [{\cal U}_z  , \ve].\ee
Hence the Fourier modes of $A_z$ transform in the adjoint of the asymptotic symmetry group, while the constant piece ${\cal U}_z$ is a connection on $S^2$. Further, \eqref{uzflatdef} and \eqref{jzaw2pt} imply that we have
\begin{equation}
\begin{split}
\<J^a_z {\cal U} (w,\bw)  O_1\cdots\>_{U=1}= \frac{T^a}{z-w} \< {\cal U} (w,\bw)  O_1\cdots\>_{U=1} . 
\end{split}
\end{equation}
A parallel structure on $\ci^-$ also exists.

The flat connection $U_z$ is related to the SCET or Wilson line fields used to study jet physics \cite{mattilya}. In CFT$_2$ with a Kac-Moody symmetry, correlations functions factorize into a 
hard part and a soft part computed by the current algebra. 4D gauge theory amplitudes also factorize into a hard and a soft part, with the latter computed by Wilson line correlators. It would 
interesting to relate this soft part to ${\cal U}$-correlators and compare it to the structure in CFT$_2$.

\section*{Acknowledgements}
We are grateful to A. Andreassen, I. Feige, S. Caron-Huot, D. Kapec, E. Kramer, V. Lysov, J. Maldacena, G. S. Ng, S. Pasterski, A. Pathak, A. Porfyriadis, M. Schwartz and A. Zhiboedov for useful conversations. This work was supported in part by DOE grant DE-FG02-91ER40654 and the Fundamental Laws Initiative at Harvard.

\appendix

\section{The soft gluon theorem}\label{appa}

In this section, we review the standard proof of the soft gluon theorem. For simplicity, we consider a theory with only scalar matter in $R_\xi$-gauge:
\begin{equation}
\begin{split}
\cl &= - \frac{1}{4g_{YM}^2} \tr{ \F_{\mu\nu} \F^{\mu\nu} } - \sum_k \left( {\cal D}_\mu \phi_k \right)^\dagger  \left( {\cal D}^\mu \phi_k \right) - \frac{1}{2\xi g_{YM}^2} \tr{ \left( \p^\mu \A_\mu \right)^2 } + \cl_{\text{gh}}  . 
\end{split}
\end{equation}
The ghost action $\cl_{\text{gh}}$ will be irrelevant at tree-level. From this action, we can determine the propagators as
\begin{equation}
\begin{split}
\bd{gluonprop}(40,30)
\fmfset{arrow_len}{2.5mm}
\fmfleft{i}
\fmfright{o}
\fmf{gluon,label=$p$,label.side=left}{i,o}
\fmfv{label=$\mu;a$}{i}
\fmfv{label=$\nu;b$}{o}
\ed  & \raisebox{4.5 mm}{ $ ~~~~~~ = \frac{- i g_{YM}^2 \delta^{ab}}{p^2 - i \e} \left[ \eta^{\mu\nu} - \left( 1 - \xi \right) \frac{p^\mu p^\nu}{p^2} \right] , $ }  ~~~~~~~~~~~~~~~~~~
\bd{scalarprop}(40,30)
\fmfset{arrow_len}{2.5mm}
\fmfleft{i}
\fmfright{o}
\fmf{fermion,label=$p$,label.side=left}{i,o}
\fmfv{label=$i$}{i}
\fmfv{label=$j$}{o}
\ed  \raisebox{4.5 mm}{ $ ~~~~~~ = \frac{-  i   \delta^{ij} } { p^2 - i \e} . $ }  
\end{split}
\end{equation}
The vertex Feynman rules are \vspace{0.3cm}
\begin{equation}
\begin{split}
\bd{3gluon}(40,40)
\fmfset{arrow_len}{2.5mm}
\fmfleft{i}
\fmfright{o1,o2}
\fmf{gluon,label=$k$,label.side=left}{g,i}
\fmf{gluon,label=$p$,label.side=left}{g,o1}
\fmf{gluon,label=$q$,label.side=left}{g,o2}
\fmfv{lab=$\mu;a$}{i}
\fmfv{lab=$\nu;b$}{o1}
\fmfv{lab=$\rho;c$}{o2}
\ed & \raisebox{6 mm}{ $ ~~~~~~~~~~~ =  \frac{1}{g_{YM}^2} f^{abc} \left[ \eta^{\mu\nu} \left( k - p \right)^\rho + \eta^{\nu \rho} \left( p - q \right)^\mu  + \eta^{\rho \mu } \left( q - k \right)^\nu \right] $ ,  } 
\\~\\
\bd{4gluon}(50,40)
\fmfset{arrow_len}{2.5mm}
\fmfleft{i1,i2}
\fmfright{o1,o2}
\fmf{gluon}{g,i1}
\fmf{gluon}{g,i2}
\fmf{gluon}{g,o1}
\fmf{gluon}{g,o2}
\fmfv{lab=$\mu;a$}{i2}
\fmfv{lab=$\sigma;d$}{i1}
\fmfv{lab=$\nu;b$}{o2}
\fmfv{lab=$\rho;c$}{o1}
\ed & \raisebox{6 mm}{ $ ~\begin{array}{c} 
 =  - \frac{i}{g_{YM}^2} \left[ f^{abe} f^{cde} \left( \eta^{\mu\rho} \eta^{\nu\sigma} - \eta^{\mu\sigma} \eta^{\nu\rho} \right) \right.   \\
  ~~~~~~~~~~~~~~ +  f^{ace} f^{bde} \left( \eta^{\mu\nu} \eta^{\rho\sigma} - \eta^{\mu\sigma} \eta^{\nu\rho} \right)  \\
  ~~~~~~~~~~~~~~~~~~~~~~~  + \left.  f^{ade} f^{bce} \left( \eta^{\mu\nu} \eta^{\rho \sigma} - \eta^{\mu \rho} \eta^{\nu \sigma} \right) \right]         , 
\end{array} $ } \\~\\
\bd{2scalar}(50,50)
\fmfset{arrow_len}{2.5mm}
\fmfleft{i}
\fmfright{o}
\fmftop{t}
\fmf{fermion,label=$k$,label.side=right}{i,g}
\fmf{fermion,label=$q$,label.side=right}{g,o}
\fmf{gluon,label=$p$,label.side=left}{g,t}
\fmfv{lab=$j$}{i}
\fmfv{lab=$i$}{o}
\fmfv{lab=$\mu;a$}{t}
\ed & \raisebox{12.5mm}{ $ ~~~~~~~~~~~  =     i \left( k^\mu + q^\mu \right) (T^a_k)_{ij}  $,  } \\
\bd{2scalar2photon}(50,40)
\fmfset{arrow_len}{2.5mm}
\fmfleft{i1,i2}
\fmfright{o1,o2}
\fmf{fermion}{i1,g}
\fmf{fermion}{g,i2}
\fmf{gluon}{g,o1}
\fmf{gluon}{o2,g}
\fmfv{lab=$\mu;a$}{o1}
\fmfv{lab=$\nu;b$}{o2}
\fmfv{lab=$j$}{i1}
\fmfv{lab=$i$}{i2}
\ed & \raisebox{6mm}{ $ ~~~~~~~~~~~  =  - i \eta^{\mu\nu}  \left( T^a_k T^b_k  \right)_{ij}$ . }
\end{split}
\end{equation}
~\\
Every external gluon is accompanied with a Feynman rule factor of $g_{YM}$.

We now consider the amplitude involving only external scalars 

\vspace{0.2cm}
\begin{equation}
\begin{split}
\bd{amp}(50,50)
\fmfset{arrow_len}{2.5mm}
\fmfleft{i1,i2,i3}
\fmfright{o1,o2,o3}
\fmf{fermion}{i1,g}
\fmf{fermion}{i2,g}
\fmf{fermion}{i3,g}
\fmf{fermion}{g,o1}
\fmf{fermion}{g,o2}
\fmf{fermion}{g,o3}
\fmfv{decor.shape=circle,decor.filled=shaded,decor.size=20}{g}
\fmfv{label=$p_1$}{o1}
\fmfv{label=$\vdots$}{o2}
\fmfv{label=$p_m$}{o3}
\fmfv{label=$\vdots$}{i2}
\fmfv{label=$p_{m+n}$}{i1}
\fmfv{label=$p_{m+1}$}{i3}
\ed
\end{split}
\end{equation}
\\
where the $k$th scalar particle is in representation $R_k$. 

We denote this amplitude as $\cm$. Now, consider the same amplitude with an additional outgoing soft gluon of momentum $p^\mu_\gamma$, color index $a$, and polarization $\e_\mu(p_\gamma)$ satisfying the gauge condition $p_\gamma \cdot \e (p_\gamma) = 0$. We denote this by $\cm^{a,\e} (p_\gamma)$. The dominant diagrams in the soft $p_\gamma^0 \to 0$ limit are 

\vspace{0.2cm}
\begin{equation}
\begin{split}
\bd{ampwithgrav}(60,60)
\fmfset{arrow_len}{2.5mm}
\fmfleft{i1,i2,i3}
\fmfright{o1,o2,o3,o5}
\fmf{fermion}{i1,g}
\fmf{fermion}{i2,g}
\fmf{fermion}{i3,g}
\fmf{fermion}{g,o1}
\fmf{fermion}{g,o2}
\fmf{fermion}{g,o3}
\fmf{gluon}{o5,g}
\fmfv{decor.shape=circle,decor.filled=shaded,decor.size=20}{g}
\fmfv{label=$p_1$}{o1}
\fmfv{label=$\vdots$}{o2}
\fmfv{label=$p_m$}{o3}
\fmfv{label=$a; p_\gamma$}{o5}
\fmfv{label=$\vdots$}{i2}
\fmfv{label=$p_{m+n}$}{i1}
\fmfv{label=$p_{m+1}$}{i3}
\ed ~~~~~~ \raisebox{28 pt}{$=~~ \sum\limits_{k=1}^m $}~~~~~~~~~
\bd{ampwithgrav1}(70,60)
\fmfset{arrow_len}{2.5mm}
\fmfleft{i1,i2,i3}
\fmfright{o1,o2,o3,o4}
\fmf{fermion}{i1,g}
\fmf{fermion}{i2,g}
\fmf{fermion}{i3,g}
\fmf{fermion}{g,o1}
\fmf{fermion}{g1,o2}
\fmf{fermion,tension=0.5}{g,g1}
\fmf{fermion}{g,o4}
\fmffreeze
\fmf{gluon}{o3,g1}
\fmfv{decor.shape=circle,decor.filled=shaded,decor.size=20}{g}
\fmfv{label=$p_1$}{o1}
\fmfv{label=$ p_k$}{o2}
\fmfv{label=$a; p_\gamma$}{o3}
\fmfv{label=$p_m$}{o4}
\fmfv{label=$\vdots$}{i2}
\fmfv{label=$p_{m+n}$}{i1}
\fmfv{label=$p_{m+1}$}{i3}
\ed  ~~\raisebox{28 pt}{$~~~~~ + ~~~  \sum\limits_{k=m+1}^{m+n}   ~ $}~~~~~~~~~
  \bd{ampwithgrav2}(70,60)
 \fmfset{arrow_len}{2.5mm}
\fmfleft{o1,o2,o4}
\fmfright{i1,i2,i3,o3}
\fmf{fermion}{g,i1}
\fmf{fermion}{g,i2}
\fmf{fermion}{g,i3}
\fmf{fermion}{o1,g}
\fmf{fermion}{o2,g1}
\fmf{fermion}{g1,g}
\fmf{fermion}{o4,g}
\fmffreeze
\fmf{gluon}{o3,g1}
\fmfv{decor.shape=circle,decor.filled=shaded,decor.size=20}{g}
\fmfv{label=$p_{m+n}$}{o1}
\fmfv{label=$p_k$}{o2}
\fmfv{label=$a; p_\gamma$}{o3}
\fmfv{label=$p_{m+1}$}{o4}
\fmfv{label=$\vdots$}{i2}
\fmfv{label=$p_1$}{i1}
\fmfv{label=$p_m$}{i3}
\ed ~~~~
\end{split}
\end{equation}
\vspace{0.2cm}

In the limit of $p_\gamma^0 \to 0^+$, we then get
\begin{equation}
\begin{split}\label{softgluon1}
\lim_{p_\gamma^0 \to 0^+} \left[ p_\gamma^0  \cm^{a,\lambda} (p_\gamma)  \right] & =  g_{YM}\left[ \sum_{k=1}^m \frac{   p_k \cdot \e_\lambda (p_\gamma) }{ p_k \cdot {\hat p}_\gamma}   T_k^a   -  \sum_{k=m+1}^{m+n} \frac{  p_k   \cdot \e_\lambda (p_\gamma) }{ p_k  \cdot {\hat p}_\gamma }  \left(   T^a_k  \right)^* \right]\cm   , 
\end{split}
\end{equation}
where $\lambda$ is the helicity of the gluon and
\begin{equation}
\begin{split}
{\hat p}_\gamma^\mu \equiv \frac{p_\gamma^\mu}{p_\gamma^0} , 
\end{split}
\end{equation}
and $T^a_k$ acts on the $k$th index on $\cm$. \eqref{softgluon1} has been derived in the context of scalar matter, but is in fact generally true for any type of matter. This is simply \eqref{tok} expressed in different notation.

We parametrize the massless momentum $p_\gamma^\mu$ in terms of $(\omega,z,\bz)$ as in \eqref{momentumparametrization} and work in a gauge where the polarization vectors take the form
\begin{equation}
\begin{split}
\e^+_\mu ( p_\gamma ) = \frac{1}{\sqrt{2}} \left( - \bz ,  1 , - i , - \bz \right) , \qquad \e^-_\mu ( p_\gamma  ) = \frac{1}{\sqrt{2}} \left( - z ,  1 ,   i , - z \right) . 
\end{split}
\end{equation}
For $\lambda = +$ and $p_k^2=0$, \eqref{softgluon1} reads
\begin{equation}
\begin{split}\label{softgluon}
\frac{1}{g_{YM}} {\hat \e}^-_z \lim_{\omega \to 0^+} \left[ \omega   \bra{ \text{out} }  \colon a_+^a \left( p_\gamma \right)^{\text{out}}   {\cal S}  \colon \ket{ \text{in} }   \right]  =  \sum_{k=1}^{m+n} \frac{ \eta_k  }{ z - z_k }  \bra{\text{out} }  : ( T_k^a   )^* {\cal S} : \ket{ \text{in} }  ,  
\end{split}
\end{equation} 
where ${\hat \e}^-_z = \frac{1}{r} \p_z x^\mu \e^-_\mu = \frac{ \sqrt{2} }{ 1 + z \bz}$ and $\eta_k = 1$ for outgoing particles and $-1$ for incoming particles. Here, we parametrized the massless momentum $p_k$ in terms of $(\omega_k,z_k,\bz_k)$. 

We now prove that this is equivalent to the soft gluon theorem. To do this, we will need to write the current $J^a_z$ in terms of creation and annihilation operators. Momentum eigenmodes in Minkowski space are usually described in flat coordinates
\begin{equation}
\begin{split}
	ds^2 = -dt^2 + d\vec{x} \cdot d \vec{x} ,
\end{split}
\end{equation}
which are related to the retarded coordinates in \eqref{retardedcoord} by
\begin{equation}
\begin{split}
	t = u + r, \quad x^1 + i x^2 = \frac{2rz}{1+z\bz}, \quad x^3 = \frac{r \left( 1 - z \bz \right)}{1 + z\bz },
\end{split}
\end{equation}
with $\vec{x} = (x^1, x^2, x^3)$ satisfying $\vec{x} \cdot \vec{x} = r^2$. 

At late times $t\to\infty$ (or equivalently $r \to \infty$ in the retarded coordinates), the gauge field $\A_\mu$ becomes free and can be approximated by the mode expansion,\footnote{In this section, we assume $C_z = 0$ (see \eqref{Czdef}) for simplicity. The final result \eqref{ndef} is independent of this assumption.}
\begin{equation}
\begin{split}\label{modexp}
	\A^a_\mu (x) = g_{YM} \sum_{\alpha=\pm} \int \frac{d^3q}{(2\pi)^3} \frac{1}{2\omega_q} \left[ \e_\mu^\alpha (q)^* a_\alpha^a (\vec{q})^{\text{out}} e^{i q \cdot x }  + \e_\mu^{\alpha} (q) a_\alpha^{a\dagger} (\vec{q})^{\text{out}} e^{- i q \cdot x }  \right] ,
\end{split}
\end{equation}
where $q^0 = \omega_q = |\vec{q}|$ and $\alpha = \pm$ are the two helicities. 

The creation and annihilation operators obey
\begin{equation}
\begin{split}
	\left[ a_\alpha^a (\vec{q})^{\text{out}}  , a_\beta^{b\dagger} (\vec{q}\,')^{\text{out}}\right] = \delta_{\alpha\beta} \delta^{ab} (2\pi)^3   (2 \omega_q) \delta^3 \left( \vec{q} - \vec{q}\,' \right) . 
\end{split}
\end{equation}
To determine $A^a_z$, recall
\begin{equation}
\begin{split}
	A^a_z (u,z,\bz) = \lim_{r \to \infty} \A^a_z(u,r,z,\bz).
\end{split}
\end{equation}
Using $\A^a_z = \p_z x^\mu \A^a_\mu$, the mode expansion in \eqref{modexp}, and the stationary phase approximation  we find
\begin{equation}
\begin{split}\label{modexp1}
	A^a_z (u,z,\bz) = - \frac{i g_{YM} }{8\pi^2}  {\hat \e}_z^- \int_0^\infty d\omega_q \left[  a_+^a ( \omega_q {\hat x} )^{\text{out}} e^{-i \omega_q u }   -  a_-^{a\dagger}( \omega_q {\hat x} )^{\text{out}}  e^{i \omega_q u }   \right]  .
\end{split}
\end{equation}
where ${\hat x}$ is defined in \eqref{xhatdef}. Using \eqref{modexp1}, we determine
\begin{equation}
\begin{split}
- i \omega A_z^{\omega,a} = - \frac{g_{YM}}{4\pi}  \hat\e^-_z  \int_0^\infty d\omega_q \omega_q \left[  a_+^a( \omega_q {\hat x} )^{\text{out}}  \delta \left( \omega - \omega_q \right)  + a_-^{a\dagger} ( \omega_q {\hat x} )^{\text{out}} \delta \left( \omega + \omega_q \right)  \right]  .
\end{split}
\end{equation}
When $\omega > 0$ ($\omega < 0$) only the first (second) term contributes. \eqref{nndeff} then gives
\begin{equation}
\begin{split}\label{ndef}
N_z^a = -  \frac{g_{YM}}{8\pi} \hat\e^-_z \lim_{\omega \to 0^+ } \left[  \omega a_+^a ( \omega {\hat x} )^{\text{out}}  + \omega a_-^{a\dagger} ( \omega {\hat x} )^{\text{out}} \right]  .
\end{split}
\end{equation}
Similarly on $\ci^-$, we find 
\begin{equation}
\begin{split}\label{mdef}
M_z^a =  - \frac{g_{YM}}{8\pi}  \hat\e^-_z \lim_{\omega \to 0^+ } \left[  \omega a_+^a ( \omega {\hat x} )^{\text{in}}  + \omega a_-^{a\dagger} ( \omega {\hat x} )^{\text{in}} \right]  ,
\end{split}
\end{equation}
where $a_\pm^{\text{in}}$ and $a_\pm^{\text{in}\dagger}$ annihilate and create incoming gluons on $\ci^-$. Crossing symmetry of \cs-matrix amplitudes implies that an outgoing positive helicity gluon has the same soft factor as an incoming negative helicity gluon up to a sign. This implies
\be\label{matching}
N_z + M_z =0,
\ee 
when inserted into scattering amplitudes. 

Similarly, insertions of $J_z^a$ are given by
\begin{equation}
\begin{split}\label{jzinstild}
\bra{\text{out}} \colon J^a_z {\cal S} \colon \ket{\text{in}}  = \frac{1}{g_{YM}} {\hat \e}^-_z  \lim_{\omega \to 0^+} \left[ \omega   \bra{ \text{out} }  \colon a_+^a \left( \omega {\hat x}(z,\bz) \right)^{\text{out}}   {\cal S}  \colon \ket{ \text{in} }   \right] . 
\end{split}
\end{equation}
Using \eqref{softgluon}, this is
\begin{equation}
\begin{split}\label{softgluon2}
 \bra{ \text{out} }  \colon J_z^a   {\cal S}  \colon \ket{ \text{in} } =  \sum_{k=1}^{m+n} \frac{ \eta_k  }{ z - z_k }  \bra{\text{out} }  : ( T_k^a   )^* {\cal S} : \ket{ \text{in} } ,
\end{split}
\end{equation} 
which reproduces \eqref{ftip}.

\bibliographystyle{utphys}

\providecommand{\href}[2]{#2}\begingroup\raggedright\endgroup

\end{document}